\documentclass{ieeeaccess}
\usepackage{cite}
\usepackage{amsmath,amssymb,amsfonts}
\usepackage{algorithmic}
\usepackage{graphicx}
\usepackage{textcomp}
\usepackage[ruled,vlined]{algorithm2e}
\def\BibTeX{{\rm B\kern-.05em{\sc i\kern-.025em b}\kern-.08em
    T\kern-.1667em\lower.7ex\hbox{E}\kern-.125emX}}
\begin{document}
\history{}
\doi{}

\title{On the experimental feasibility of quantum state reconstruction via machine learning}
\author{\uppercase{Sanjaya Lohani}\authorrefmark{1,3},
\uppercase{Thomas A. Searles\authorrefmark{1,2}, Brian T. Kirby\authorrefmark{3,4}, and Ryan T. Glasser}\authorrefmark{3}}
\address[1]{IBM-HBCU Quantum Center, Howard University, Washington, DC 20059, USA (email TAS: thomas.searles@Howard.edu)}
\address[2]{Massachusetts Institute of Technology, Cambridge, MA 02139, USA}
\address[3]{Tulane University, New Orleans, LA 70118, USA (email RTG: rglasser@tulane.edu)} 
\address[4]{United States Army Research Laboratory, Adelphi, MD 20783, USA (email BTK: brian.t.kirby4.civ@mail.mil)}
\tfootnote{This material is based upon work supported by, or in part by, the Army Research Laboratory and the Army Research Office under contract/grant numbers W911NF-19-2-0087 and W911NF-20-2-0168. T.A.S. acknowledges support from the IBM-HBCU Quantum Center and the Martin Luther King Visiting Scholars Program at MIT.
}

\markboth
{Lohani \headeretal: ON THE EXPERIMENTAL FEASIBILITY OF QUANTUM STATE RECONSTRUCTION VIA MACHINE LEARNING}
{Lohani \headeretal: ON THE EXPERIMENTAL FEASIBILITY OF QUANTUM STATE RECONSTRUCTION VIA MACHINE LEARNING}

\corresp{Corresponding author: Sanjaya Lohani (email: sanjaya.lohani@Howard.edu).}

\begin{abstract}
We determine the resource scaling of machine learning-based quantum state reconstruction methods, in terms of inference and training, for systems of up to four qubits when constrained to pure states.  Further, we examine system performance in the low-count regime, likely to be encountered in the tomography of high-dimensional systems.  Finally, we implement our quantum state reconstruction method on an IBM Q quantum computer, and compare against both unconstrained and constrained MLE state reconstruction.
\end{abstract}

\begin{keywords}
Machine Learning, Quantum Tomography, IBM Q
\end{keywords}

\titlepgskip=-15pt

\maketitle

\section{Introduction}
\label{sec:introduction}
\PARstart{Q}{uantum} state tomography is a standard procedure for determining the state of an unknown quantum system through a series of repeated measurements on an ensemble of identically prepared systems.  The number of measurement settings required for full state tomography of a multi-qubit state scales exponentially with the qubit number\cite{haffner2005scalable,vrehavcek2004minimal,wootters1989optimal,altepeter2005photonic}.  Further, the determination of a physically valid density matrix that corresponds to the measured data is itself resource-intensive\cite{hou_full_2016,smolin_efficient_2012}.  Many methods for reconstructing valid density matrices from measured data have been developed, such as Bayesian \cite{lukens2020practical}, maximum likelihood estimation (MLE)\cite{teo2011quantum,james2005measurement,smolin_efficient_2012}, projected gradient descent\cite{bolduc2017projected}, and linear regression\cite{qi2017adaptive}.  
Recently, several methods for state reconstruction based on machine learning have been proposed and demonstrated \cite{carrasquilla2019reconstructing,cha2020attention,tiunov2020experimental,torlai2019integrating,palmieri2020experimental,torlai2018neural,neugebauer2020neural,lohani2020machine,xu2018neural}.  These machine learning-based techniques offer several potential advantages compared to their conventional counterparts, such as the ability to front-load expensive computations, and robustness to missing measurements and experimental noise\cite{danaci2020machine}.  Despite these advantages, it is not yet clear how a machine learning models' training scales with system size, nor how they perform in situations where limited copies of the unknown state are available.  

In this paper, we implement machine-learning-based quantum state reconstruction for systems of up to four qubits where states are assumed pure.  
{This assumption differs from our previous work reconstructing mixed states \cite{lohani2020machine,danaci2020machine}. 
It is motivated by an effort to achieve both the highest possible reconstruction fidelity and the most favorable resource scaling by restricting the size of the accessible Hilbert space. 
Such an assumption is potentially applicable in quantum computing, where systems under investigation can be localized, as opposed to quantum communication, where channel impairments are unavoidable and quickly decohere transmitted states.}
Based on simulation with Haar random pure states, we find that our networks reconstruct states of up to four qubits comparably to an MLE process restricted to pure states but with a higher variance.  
Based on our results, we estimate that training time increases exponentially with qubit number, but that inference time remains linear.
In other words, we find that despite a steep up-front training cost, we can achieve relatively similar reconstruction fidelity as MLE but significantly faster once trained.
These results can be summarized as a trade-off between rapid and high average reconstruction fidelity with our machine-learning-based methods and relatively slow but lower variance reconstruction with MLE.

\Figure[h!](topskip=0pt, botskip=0pt, midskip=0pt)[width=\linewidth]{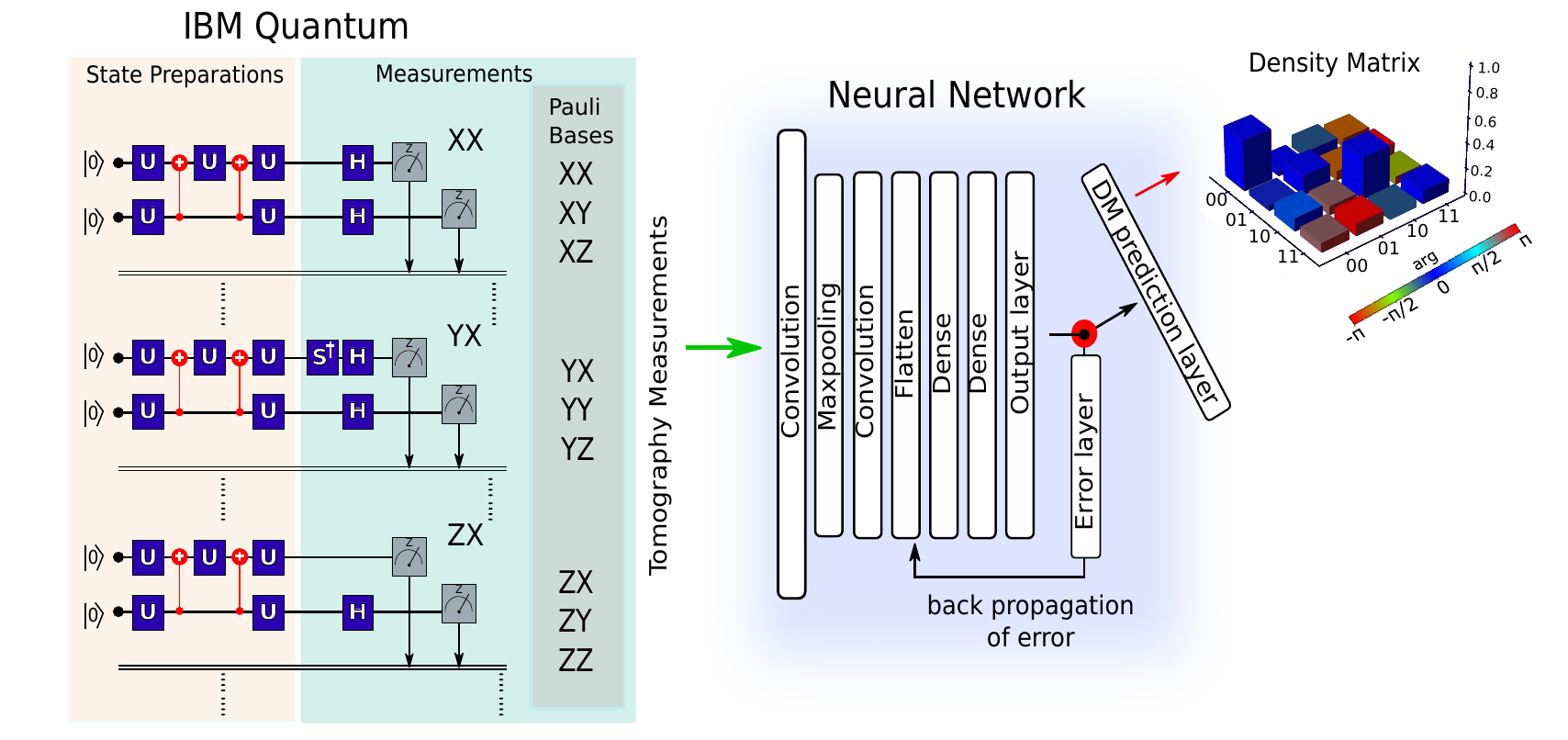}
{{\textbf{Example of an end-to-end tomography setup for two qubits.}  The $X$, $Y$, and $Z$ represent the Pauli basis, the circuits for which are prepared consecutively. The neural network then uses the measurement results to reconstruct a density matrix, as illustrated on the right.}
\label{fig:Figure_1}}

{We then investigate the impact of statistical noise on our neural network system by limiting how many copies of an unknown state are available for tomographic measurements. 
The generative adversarial learning technique explored in \cite{ahmed2020quantum, ahmed2020classification} uses a {randomly initialized network that is tuned based on experimental data in order to make make state estimations.} 
The added parameter tuning of these systems {based on actual data taken from the state to be reconstructed} requires further iterations, increasing the overall computational cost. 
Alternatively, in this paper, we explore the approach of explicitly training our network on simulated data that already includes statistical noise, meaning we artificially restrict the number of copies used to build measurement statistics in the training set.
Surprisingly, we find that a network trained with data that includes statistical noise (restricted number of copies used to build measurement statistics) outperforms a network trained on ideal data (infinite copies used to build measurement statistics) when reconstructing experimental data that itself includes statistical noise.
We conclude that the variations in measurement results due to limited counting statistics are themselves learnable, to some degree, by our network. 
These results inform the applicability of neural network quantum state reconstruction for high-dimensional systems where resource limitations in computation and the repeatability of measurements are important factors. }

Finally, we then use our trained networks to reconstruct states generated and measured on the IBM Q quantum processor and compare against several other techniques.  
The limited quantum volume of the IBM Q quickly makes our assumption of purity invalid.  We then determine the performance of such a special-purpose system, meaning state reconstruction methods trained only with pure states, in a general environment by comparison against unconstrained Gaussian {and multinomial} MLE reconstruction. 
{We find that our network trained exclusively with pure states performs comparably to constrained MLE when used to reconstruct mixed states.
In particular, we note that the performance of the constrained MLE and pure-state trained network approximately tracks the purity of the system under investigation.
This is somewhat surprising as our network is not constrained to reconstruct only pure states, rather, it was merely not exposed to them during training.
These results indicate the difficulty a neural-network based tomography system has reconstructing states which significantly differ from the training set and signal the caution required when deploying a pre-trained system in diverse environments.}

\section{Methods}
\subsection{Generating random quantum states and simulating state tomography}
\label{subsec:generating}
{To generate random pure quantum states, $|\psi\rangle$, we first expand them as
\begin{equation}
   |\psi\rangle\, = \, \sum_{i=1}^{2^d} u_i |\phi\rangle_i, 
   \label{eqn:pure_state}
\end{equation}
where `d' represents the number of qubits in the state, $u_i$ represents the $i^{th}$ element of the first column of a $2^d\times2^d$ random unitary matrix, and the $|\phi_{i}\rangle$ represent computational basis states of appropriate dimension \cite{bengtsson2017geometry}. 
For example, for two qubits, $|\phi\rangle\,=\,\{ |00\rangle,\, |01\rangle,\, |10\rangle,\, |11\rangle \}$. 
Furthermore, to avoid possible convergence issues with the Cholesky decomposition of a pure density matrix \cite{lohani2020machine,higham_analysis_1990}, we add a small perturbation term, $\epsilon\,=\,1\times10^{-7}$, to the generated states as given by,
\begin{equation}
   \rho\,=\, (1-\epsilon)|\psi\rangle\langle\psi| + \frac{\epsilon}{4}I.
    \label{eqn:epsilon}
\end{equation}
For each $\rho$ we then apply the Cholesky decomposition $\rho\,=\,TT^{\dagger}$ where $T$ is a lower triangular matrix \cite{higham_analysis_1990}, and collect the elements of $T$ into a $\tau$-vector given by [$\tau_0$, $\tau_1$, $\tau_2$, $\tau_3$, .. .. .., $\tau_{2^{2d}-1}$].
As an example, in the two-qubit case the $T$ matrices of the Cholesky decomposition are given by
\begin{equation}
\begin{aligned}
    &T\,=\,
        \begin{bmatrix}
        \tau_0 & 0& 0& 0\\
        \tau_4+i\tau_5 & \tau_1 &0 &0\\
        \tau_{10}+i\tau_{11} & \tau_6+i\tau_7 &\tau_2 &0\\
        \tau_{14}+i\tau_{15} & \tau_{12}+i\tau_{13} &\tau_8+i\tau_9 &\tau_3\\
        \end{bmatrix}, \\
\end{aligned}
\label{eqn:lower_t}
\end{equation}
with corresponding $\tau$-vectors 
\begin{equation}
\begin{aligned}
\quad T \rightarrow \tau = [\tau_0, \, \tau_1,\, \tau_2,\, \tau_3,\,  .. .. ..,\, \tau_{15}].
\end{aligned}
\label{eqn:lower_T_to_t}
\end{equation}
In the ideal case, where we assume we have infinite copies of an unknown state to build measurement statistics with, tomographic measurement results are calculated directly from the expectation value of each measurement operator.
More precisely, these probabilities are calculated from
\begin{equation}
    \bar{n}_{i}\,=\, \textrm{Tr}\big(\rho\mathcal{\hat{\varPi}}_i\big).
    \label{eq:measurement}
\end{equation}
{where `Tr' represents the trace, and $\mathcal{\hat{\varPi}}$ is the set of $6^{d}$ projectors created from the tensor product of the eigenvectors of all $3^{d}$ combinations of Pauli operators.}
Hence, the index $i$ in (\ref{eq:measurement}) runs from $1$ to $6^d$.
{For example, for a single qubit, $\mathcal{\hat{\varPi}}$ consists of the six eigenvectors associated with $\{X,Y,Z\}$ where $X$, $Y$, and $Z$ are the usual Pauli operators.
For $d$ qubits this extends naturally with $\mathcal{\hat{\varPi}}$ containing the $6^{d}$ eigenvectors associated with the $3^{d}$ combinations of operators $\{X,Y,Z\}_{1}\otimes\{X,Y,Z\}_{2}\otimes...\otimes\{X,Y,Z\}_{d}$ where the subscripts indicate qubit number.}
While this choice of measurement procedure is over-complete \cite{vrehavcek2004minimal}, it is often performed in practice due to its relative simplicity and results showing such a basis set minimizes statistical error \cite{wootters1989optimal}.}

{
We also consider the situation where a limited number of copies of a quantum state are available for tomographic measurement, meaning measured probabilities will vary statistically from the expectation values, and we can no longer use the procedure outlined above.
To generate data sets with the appropriate statistical noise, we follow the procedure outlined in Appendix \ref{app:low}.}

\subsection{Training the network}
{We build a custom-designed convolutional neural network with a convolutional unit of kernel size (2, 2), strides of 1, ReLU as an activation function, and 25 filters.  The filters are followed by a max-pooling layer with pool-size (2, 2) and a second convolutional unit with the same configuration in a row. We then connect two dense layers (Dense 1 and Dense 2), each followed by a dropout layer with a rate of 0.5, which is finally attached to an output layer that predicts the $\tau$-vectors. Then, the predicted $\tau$-vectors are forwarded to an error layer, where mean square loss between the target and predicted $\tau$ is evaluated and fed back to optimize the network's training using the Adagrad optimizer with a learning rate of 0.005 for up to 300 epochs. Additionally, the output layer is also branched to a DM-prediction layer as shown in Fig. \ref{fig:Figure_1}, where predicted $\tau$-vectors are further re-arranged to lower triangular matrices, $T$, as expressed in (\ref{eqn:lower_t}), and corresponding density matrices are predicted as, $\rho_{nn}=\frac{TT^\dagger}{\textrm{Tr}(TT^\dagger)}$. Prediction of $T$ rather than $\rho$ directly ensures that the predicted density matrices from the network are always physical \cite{altepeter2005photonic}}. Also, note that there are no training parameters between the error layer and the density matrix (DM) prediction layer. {We build the DM prediction layer in the same graph of the network to efficiently evaluate the fidelity metric per epoch, to cross-validate the training, and to avoid post-processing.  In order to evaluate the fidelity ($F$) between the predicted density matrix ($\rho_{nn}$) and the target ($\rho_t$), we use $F = \Big|\textrm{Tr}\sqrt{\sqrt{\rho_{nn}}\rho_{t}\sqrt{\rho_{nn}}}\Big|^2$. 
The input size, output size, and number of the fully connected dense layers, and the number of fully-connect neurons, as a function of qubit number are shown in Table \ref{table:one}. 
Note that we keep the depth (number of hidden layers) of the network fixed for all cases; instead, we increase the number of the neurons in the dense layers with increasing qubit number.
\begin{table}[h!]
\centering
 \begin{tabular}{|p{1cm}|p{1cm}|p{1cm}|p{1cm}|p{1cm}|} 
 \hline
 \# Qubit & Input & Dense 1 & Dense 2 & Output \\ [1ex] 
 \hline\hline
 1 & [2, 3] & 250 & 150 & 4  \\ 
 2 & [6, 6] & 750 & 450 & 16\\
 3 & [6, 36] & 2500 & 1000 & 64 \\
 4 & [36, 36] & 4500 &  2500 & 256 \\ [1ex] 
 \hline
 \end{tabular}
\caption{Number of neurons in fully-connected layers as a function of the number of qubits in the state.}
\label{table:one}
\end{table}
}
{The number of quantum states used for training is manually optimized as a function of qubit number. 
The parameters to be learnt while training the network are termed as trainable parameters. The number of trainable parameters increases (decreases) with increasing (decreasing) number of fully-connected neurons and input size. Further, with respect to reconstruction fidelity, computational cost, and generality, we find an optimized number of quantum states in a training set to be 35,000 for all cases. As an explicit example, we present the reconstruction fidelity with varying trainable parameters and number of quantum states in the training set for the case of four qubits in Appendix \ref{app:par} (Fig. \ref{fig:appendix-1}).} 

{To train the network, we generate 35,500 random quantum states according to the Haar measure as described by (\ref{eqn:pure_state}) and the associated $6^d$ tomographic measurement for systems with qubit numbers ranging from one to four.
We split the simulated data into a training set of size 35,000 and a validation set of size 500 to cross-validate the network performance. } To reduce the bias-variance trade-off and increase the fidelity at the output, we use separate networks (see Table \ref{table:one}) for different qubit numbers. Note that we implement a batch size of 100 in the training of each network. After training, we generate test sets that are entirely unknown to the trained network. {For the test-cases, we consider two scenarios, a simulated test-set without any device noise and a test set directly measured from a near-term intermediate scale quantum computer with {7 qubits (IBM Quantum's $ibmq\_jakarta$)}.}



\Figure[h!](topskip=0pt, botskip=0pt, midskip=0pt)[width=\linewidth]{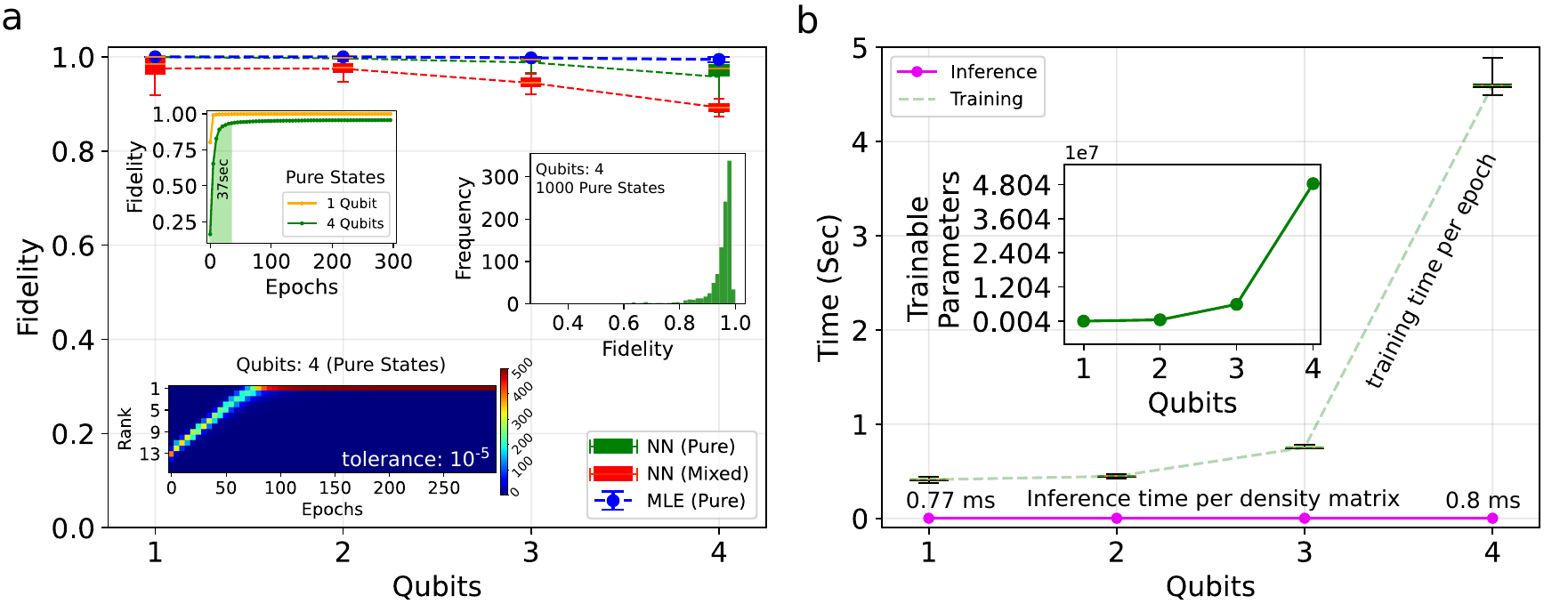}
{{\textbf{Reconstruction fidelity and resource scaling}. (a) Fidelity for the predicted density matrices versus the number of qubits. The blue dashed line shows the reconstruction fidelities using maximum likelihood estimation (MLE) with multinomial likelihood constrained to reconstruct only pure-states, the green box represents the reconstruction fidelities using a neural network (NN) {trained with pure states predicting pure states, whereas the red box shows the reconstruction fidelities using a neural network trained with mixed states predicting mixed states}. The top-left inset shows the cross-validation fidelity for up to 300 epochs in one and four qubit tomography {for pure states}, where the bottom index shows the rank of the predicted quantum states per epoch, {again for pure states}. {Similarly, the top-right inset shows a histogram of the reconstruction fidelity for 1000 random pure states of four qubits using our neural network.} (b) Network inference times versus the number of qubits. {In each plot the box represents the interquartile range and the whiskers are at the $5^{\text{th}}$ and $95^{\text{th}}$ percentiles.}
}
\label{fig:Figure_2}}

\subsection{Maximum likelihood estimation (MLE)}
{For comparison, we have also implemented multiple maximum likelihood reconstruction techniques.
We begin by discussing the Gaussian likelihood approach. 
As described in Section \ref{subsec:generating}, for the tomography method we employ there are $6^d$ total measurement outcomes $n_i$ for a $d$ qubit state. 
The expected value of each measurement outcome is denoted by $\bar{n}_{i}$ and are again given by (\ref{eq:measurement}). 
For simplicity, we first assume Gaussian noise, which leads to the probability ($P$) of obtaining the outcome $n_i$ given by
\begin{equation}
    P_{G}(n_1, n_2,..., n_{6^d})\,=\,\prod_{i=1}^{6^d} \exp{\frac{-(n_i - \bar{n}_{i})^2}{2\sigma_i^2}}.
\end{equation}
Here $\sigma_i$ is approximated as $\sqrt{\bar{n}_{i}}$.
Using (\ref{eq:measurement}), we find the negative log-likelihood of the given density matrix $\rho$ is given by, 
\begin{equation}
    -\mathcal{L} = \sum_{i=1}^{6^d}\frac{\Big[\textrm{Tr}\big[\rho\mathcal{\hat{\varPi}}_i\big]-n_i\Big]^2}{2\textrm{Tr}\big[\rho\mathcal{\hat{\varPi}}_i\big]}.
\label{eqn:log-likelihood}
\end{equation}}

{Next, we implement the multinomial likelihood approach. For this, we assume the joint conditional probability of a set of measurement results on $\rho$ expressed as,
\begin{equation}
    P_{M}(n_1, n_2,..., n_{6^d},N_1,N_2,...,N_{6^d})\,=\,\prod_{i=1}^{6^d}\textrm{Tr}(\rho\mathcal{\hat{\varPi}}_i)^{N_i},
\label{eqn:multi-prob}
\end{equation}
where $N$ is the number of times the given projector ($\varPi$) has been executed. Instead of directly maximizing (\ref{eqn:multi-prob}), we define a negative log-likelihood as given below, and minimize it:
\begin{equation}
    -\mathcal{L}\,=\,-\sum_{i=1}^{6^d} N_i log\big[\textrm{Tr}(\rho\mathcal{\hat{\varPi}}_i)\big].
\label{eqn:multi-prob-likelihood}
\end{equation}
}
{In order to optimize the likelihood, we use a constrained and unconstrained optimization as discussed below.}

{In unconstrained optimization we allow for the reconstruction of any physical density matrix, including the mixed states.
As described above, with no loss of generality we can enforce the physical nature of a reconstructed density matrix by reconstructing the $\tau$-vectors instead of the density matrix directly. 
Note that the size of the $\tau$-vectors scales as $2^{2d}$ and that in this framework any density matrix can be written as
\begin{equation}
\resizebox{0.5\textwidth}{!}{$
\begin{aligned}
    & \rho(\tau_0, \tau_1, ... \tau_{2^{2d}-1})\,=\,\frac{TT^{\dagger}}{\textrm{Tr}(TT^{\dagger})},\\
    & \textrm{as},\, T\,=\,
        \begin{bmatrix}
        \tau_0 & 0& 0 & .. &0 \\
        \tau_{2^{d}}+i\tau_{2^{d}+1} & \tau_1 &0 & .. & 0 \\
        \tau_{3(2^{d}-1)+1}+i\tau_{3(2^{d}-1)+2} & \tau_{2^{d}+2}+i\tau_{2^{d}+3} &\tau_2 & .. & 0 \\
        .. & ..& .. & .. & 0 \\
        .. & ..& .. & \tau_{3(2^{d}-1)-1}+i\tau_{3(2^{d}-1)}& \tau_{2^{d}-1}\\
        \end{bmatrix}, \\
\end{aligned}$}
\label{eqn:mle_uncons_gaussian_rho}
\end{equation}
which we use to optimize over.
}
{Alternatively, for a fair comparison to our neural network that has been trained only on pure states, we also consider an optimization which is constrained to predict only pure states.
Here we only need to optimize over $2^{d+1} - 1$ unknown coefficients, $\{u\}$, analogous to (\ref{eqn:pure_state}), such that,
\begin{equation}
    \rho (u_0, u_1, u_2, ..., u_{2^{d+1} - 2}) = \frac{|\psi\rangle\langle\psi|}{\textrm{Tr}(|\psi\rangle\langle\psi|).}
\label{eqn:psi-psi-rho}
\end{equation}}

{
In both cases, constrained and unconstrained, we maximize the likelihood by minimizing the right-hand side of (\ref{eqn:log-likelihood}, \ref{eqn:multi-prob-likelihood}) using the BFGS optimizer of the tensorflow-probability library. Note that we always randomly initialize the values for $[\tau_0, \tau_1, \tau_2, ..., \tau_{2^{2d}-1}]$ and $[u_0, u_1, u_2, ..., u_{2^{d+1} - 2}]$  using the Glorot-Uniform initializer of the tensorflow-keras library, and implement a gradient tolerance of $10^{-5}$. Additionally, instead of evaluating a good initial point \cite{usami2003accuracy, james2005measurement} to avoid local convergence\cite{mascarenhas2004bfgs}, we run MLE with 50 random initial points for each quantum state and choose the state with the maximum reconstruction fidelity.}

\section{Results and Discussion}
\subsection{Resource scaling}
An illustrative example of an end-to-end tomography setup with a neural network is shown in Fig. \ref{fig:Figure_1}. We implement the network using tensorflow\cite{tensorflow2015-whitepaper} with an AMD Ryzen 9 3900x 12-core processor and GeForce RTX 2060 GPU. 
First, we make predictions for 50 test quantum states using a network solely trained with a simulated tomography data set for 300 epochs. 
To simulate the data we consider measurements on an infinitely large ensemble of identical quantum states, meaning that statistical noise should approach zero, as described in Section \ref{subsec:generating}. 
For clarity we refer to such data sets as `ideal tomography data' in the paper. Note that the test set is completely unknown to the network. {In order to illustrate the network's efficacy, we compare the network performance with MLE constrained to reconstruct pure states}. 
The results are shown in Fig. \ref{fig:Figure_2}.

{The fidelity between the reconstructed states using a neural network trained with pure states} and the corresponding target density matrices for systems of one, two, three, and four qubits are shown in Fig. \ref{fig:Figure_2} (a) by green box plots. The whisker on the box represents a fidelity ranging from {$5^{th}$ to $95^{th}$} percentiles. {Note that the results shown in the {top-left and bottom insets} are for {pure} quantum states from the validation set as described in ``Methods''.  We see from Fig. \ref{fig:Figure_2} (a) that our neural network trained on only pure states reconstructs with an average fidelity near unity, but with significantly more variance than constrained MLE.} 
The reconstruction fidelity per epoch for random quantum states in the cases of one and four qubits is shown in the top{-left} inset. We find that the training of the network takes only 1.22 seconds for a single qubit, and 37 seconds for four qubits in order to reach saturation. Additionally, we track the rank of the density matrices predicted by the network. 
In order to evaluate the rank, we use a tolerance of $1\times10^{-5}$. 
A rank of 1 corresponds to a pure-state, otherwise the predicted matrices are mixed states. As an illustration, we include an inset (bottom) showing the rank of predicted quantum states per epoch in the case of 4 qubits. 
The colorbar on the right represents the number of quantum states totalling to 500 from the validation set. 
{We find that our network trained with only pure states efficiently learns to reconstruct states with low rank despite the absence of a formal constraint requiring this, such as in constrained MLE.} {Similarly, we plot a histogram showing the reconstruction fidelities for 1000 random four qubit pure states. We find that most states, $89\%$, are well above a fidelity of $0.9$ and only a tiny fraction, $3.8\%$, of states are below $0.8$, as shown in the top-right inset.} 
{Next, we generate mixed states as described in \cite{lohani2020machine}, and use them to train and validate a separate neural network for each qubit case.}
{In particular, the network is trained and tested using randomly generated states sampled from the Hilbert-Schmidt distribution.}
{Note that the architecture and hyper-parameters of the mixed-state network and the size of training, validation, and test sets are the same as previously mentioned for the pure state network. Once the network is trained, it reconstructs the mixed states from the test set. The reconstruction fidelity is shown by the red box in Fig. \ref{fig:Figure_2} (a).}
{We see that, for a given set of network parameters, the restriction of the Hilbert space to pure states alone (green) outperforms the network trained and tested over the entire Hilbert space (red).
Hence, a restriction of the accessible Hilbert space of a machine-learning based tomography system appears to allow for an improved performance on the set of target states.
We stress that in this demonstration, we have kept the parameters and architecture of the networks equivalent to emphasize the role of the size of the Hilbert space on the reconstruction fidelity. However, improvement of the performance of the mixed-state (red) network is potentially possible at the cost of worse resource scaling, i.e., a network with more trainable parameters and more time to train. }
{Unless explicitly stated we will use neural networks trained with only pure states for the remainder of the text in order to achieve the most favorable scaling possible.}

Also in \ref{fig:Figure_2} (a) we reconstruct the same 50 test quantum states using MLE with a multinomial likelihood constrained to only pure-states (see ``Methods''), with the corresponding fidelities shown by the blue dashed line. {We find the {$5^{th}$ to $95^{th}$} percentiles for the reconstructed fidelities ranges from 0.992 to 0.999, and 0.984 to 0.999, respectively, corresponding to the case of 3 and 4 qubits.}


Next, we evaluate the computational cost of training and implementing our network. As the number of qubits, `d', increases, the number of quantum circuits required for full state tomography becomes $3^d$ and the number of possible measurement outcomes becomes $6^d$. Because of this we increase the network training parameters accordingly in order to achieve similar fidelities for systems of all qubit numbers {as expressed in Table \ref{table:one}}. The average training time per epoch versus number of qubits is shown by the green dotted line in Fig. \ref{fig:Figure_2} (b). The box plots represent the time per epoch in the {$5^{th}$ to $95^{th}$} percentiles. 
As expected we find that the training time increases with the number of qubits in the state. The sudden rise in the training time for systems of size four is due to a drastic increase in the trainable parameters as shown in the inset. {To further illustrate the effect of the number of trainable parameters on the overall fidelity we present a plot of fidelity versus trainable parameters for the four qubit case in Appendix \ref{app:par} (Fig. \ref{fig:appendix-1})}. Once the network is trained, we find that the network makes the prediction of a density matrix for system sizes of one and four qubits, respectively, within 0.77 milliseconds and 0.8 milliseconds. 

\subsection{Performance in the low count regime}
\Figure[h!](topskip=0pt, botskip=10pt, midskip=0pt)[width=.99\linewidth]{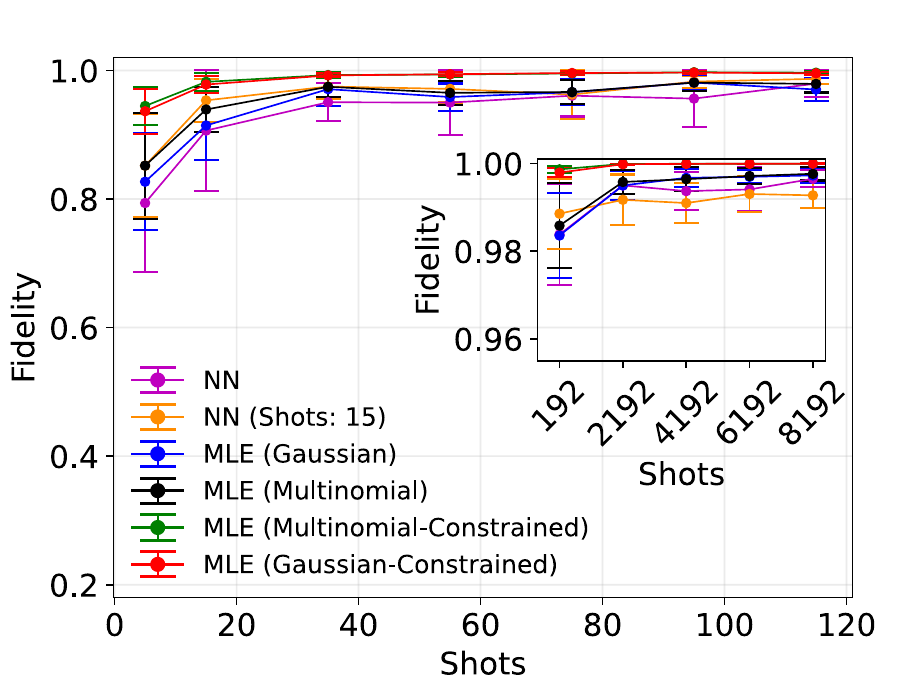}
{{\textbf{State reconstruction as a function of shots for two-qubit states.} The inset shows the limit of a large number of measurements, corresponding to minimum statistical noise. {The error bars represent one standard deviation from the mean.}} \label{fig:Figure 3}}

We now investigate the performance of our quantum state reconstruction method in the scenario that only a limited number of copies of a state are available for measurement.
Hence, we are assuming that the $6^d$ measurement results differ significantly from those predicted by (\ref{eq:measurement}) due to statistical noise.  
In this section we use the terminology `shot' to indicate the number of times each measurement circuit runs. 
In other words, the number of times we obtain a complete set of tomography for the given state.  {For example, in the two-qubit case we use 9 measurement circuits resulting in $36$ possible measurement outcomes, and hence $n$ shots would indicate we have executed all 9 circuits $n$ times.} 

In order to illustrate the proof of concept, we fix the system size at two qubits and vary the number of shots when using different reconstruction methods, plotting the results in Fig. \ref{fig:Figure 3}.
Here we have implemented {six methods: a network trained with ideal tomography data ``NN'', a network trained with the tomography data simulated at 15 shots ``NN (Shots: 15)'', MLE with a Gaussian likelihood, MLE with a multinomial likelihood, MLE with a Guassian likelihood but constrained to only pure states, and finally MLE with a multinomial likelihood but constrained to only pure states.
Details about how we generate data sets with statistical noise can be found in Appendix. For each of the listed methods, we simulate tomography for shot numbers ranging from 5 to 8192 for 20 random quantum states.} The average reconstructed fidelity at various shots is shown in Fig. \ref{fig:Figure 3}, with the fidelity in the  higher shot regime shown in the inset. 
As expected, we find that the network trained with ideal tomography data {(magenta line)} works better at higher shot numbers and reaches a fidelity of 0.997 at 8192 shots. Additionally, we show the fidelity at 5 and 15 shots are enhanced with the network that is trained with the simulated data at 15 shots (an {orange} line). Furthermore, in order to illustrate the performance efficacy of the networks, we also reconstruct random quantum states from the same test-set with multiple maximum likelihood methods as shown in Fig. \ref{fig:Figure 3}. {The error bars on the plot show one standard deviation from the mean}.

{We find that the network trained with ideal tomography data achieves a reconstruction fidelity above $0.99$ in the high-shot regime.}
Notably, the network trained for simulated tomography of 15 shots produces higher average fidelity than the network trained with ideal tomography data and the MLE with Gaussian likelihood (blue line). 
Further, the 15 shot neural network method is effectively equivalent to MLE with a multinomial-likelihood ({black} line) in the lower shot regime.
Therefore, we conclude that machine-learning-based quantum state reconstruction methods should consider the expected number of repeated measurements during training. 
This result has implications for the reconstruction of systems with high qubit numbers, where system constraints likely keep the number of repeated measurements low.

\subsection{Comparison of methods using IBM Q} 
\Figure[h!](topskip=0pt, botskip=10pt, midskip=0pt)[width=.99\linewidth]{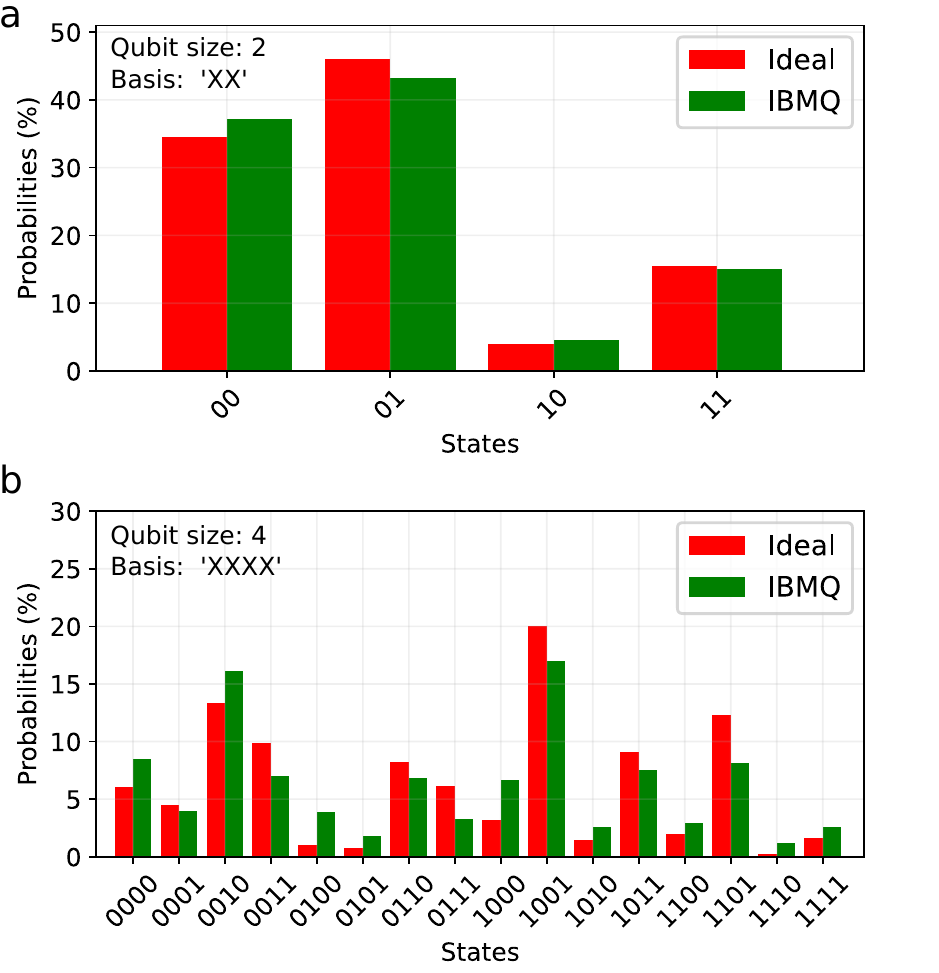}
{{\textbf{Tomography from the real IBM Q computer at {8000} shots versus ideal (simulated) tomography data}. Examples of (a) 2 qubits-tomography measured in the `XX' basis, (b) 4 qubits-tomography measured, again, in the `XXXX' basis. } \label{fig:Figure_4}}
{We now apply our neural network reconstruction technique to tomography data obtained from an IBM Quantum (IBM Q) computer and compare the results with various MLE implementations.} {We begin by implementing a $random\_circuit$ module from the Qiskit Terra API \cite{Qiskit} to generate 20 random quantum circuits for systems ranging from one to four qubits. Note that in each case, the qubits are initialized in the $|0\rangle$ state. We then upload the circuits to the IBM Q hardware, $ibmq\_jakarta$. The uploaded circuits are further transpiled by the backend. We find the minimum and maximum depths (including measurements gates) of 2 and 9, 6 and 22, 8 and 34, and 12 and 49, respectively, for the transpiled quantum circuits for systems of one through four qubits. We then execute the transpiled circuits for 8000 shots and record the measurement counts. Finally, we implement multinomial likelihood estimation to estimate the state of the system.}

{To demonstrate the extent of the noise introduced by the IBM Q components during tomography, we show with green bars in Figs. \ref{fig:Figure_4} (a) and (b) example measurement results for random two and four qubit quantum states measured in the `XX' and {`XXXX'} basis respectively.  Each circuit generating Figs. \ref{fig:Figure_4} (a) and (b) is executed for 8000 shots.
We also simulate the ideal projective measurement probabilities for the same states and show these in red on the same plots. 
The y-axis probabilities are expressed in percentage and are calculated as a ratio of the total counts to the total shots. 
We find that the depth of the self-transpiled quantum circuits for random quantum states increases with qubit number. 
As a result, we obtain more noise in the tomography for higher qubits. 
This increasing noise with qubit number is illustrated in Fig. \ref{fig:Figure_4} (b), where we show a significant difference in the measurement results between simulated and measured for four qubits, which is accompanied by a larger self-transpiled quantum circuit by the backend, {$ibmq\_jakarta$}.
{Finally, we note that an additional source of error that likely causes a mismatch between the measured and simulated results in Fig. \ref{fig:Figure_4} is how accurately the IBM Q implements the measurement projectors, which are assumed perfect in simulation.}}

\Figure[h!](topskip=0pt, botskip=10pt, midskip=0pt)[width=.99\linewidth]{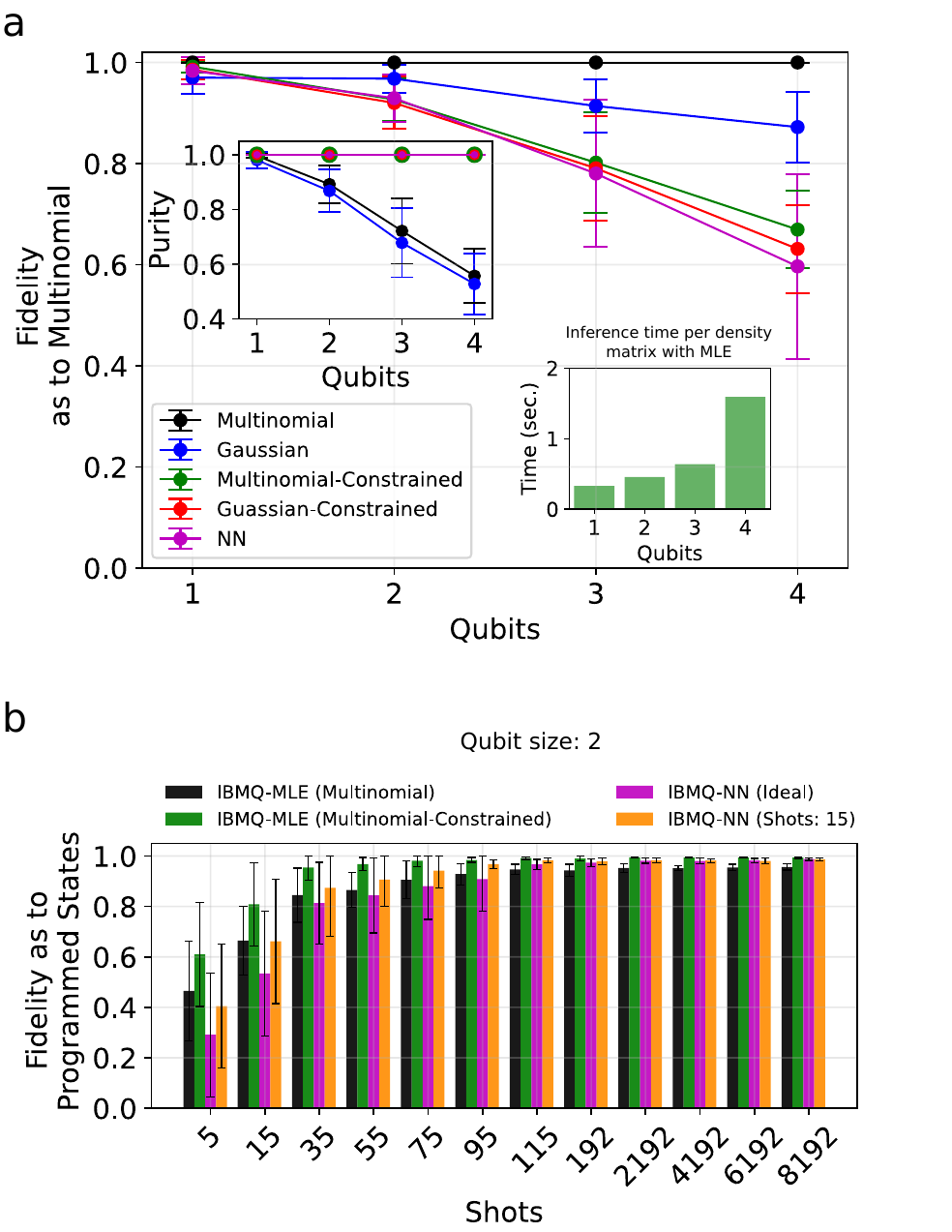}
{{\textbf{Reconstructing quantum states using measurements collected from the IBM Q.} (a) Reconstruction fidelity {with multinomial MLE as the reference} versus the number of qubits using measurement data obtained from the IBM Q at a shot value of {8000}. {The purity of states reconstructed is shown in the top-inset}.  Inference time per density matrix with maximum likelihood estimation is presented in the {bottom-right} inset.  This can be contrasted with the inference time for our neural network technique ($\approx$0.8 ms for four qubits) in Fig. \ref{fig:Figure_2} (b). (b) Reconstruction fidelity {as calculated against the state programmed into the IBM Q} for two-qubit states as a function of shots}.  {In each figure the error bars represent one standard deviation from the mean.  } \label{fig:Figure_5}}

{We now perform state reconstruction using several methods using measurement results obtained from the IBM Q. The fidelities of the reconstructed states against those obtained from the multinomial likelihood method as a function of qubit number are shown in Fig. \ref{fig:Figure_5} (a).  Here the blue, green, red, and magenta lines, respectively, correspond to MLE with a Gaussian likelihood, MLE with multinomial likelihood constrained to only pure-states, MLE with Gaussian likelihood constrained to only pure-states, and our neural network trained exclusively on pure states. For perspective, we have also included the black line, which shows the fidelity of states reconstructed using MLE with a multinomial likelihood.  Since, as mentioned above, MLE with a multinomial likelihood is our reference state for calculating fidelity, the black line is one in all cases. We also evaluate the purity of the states reconstructed with each method and show the results in the upper inset.  Finally, we note that the errors bars in Fig. \ref{fig:Figure_5} (a) represent one standard deviation from the mean.}

{Interpretation of the results of Fig. \ref{fig:Figure_5} (a) must take into account the discrepancy between the state programmed into the IBM Q and the state generated, as emphasized by Fig. \ref{fig:Figure_4}.
Since Fig. \ref{fig:Figure_4} suggests mixing of the state by the IBM Q, it is reasonable to assume that the unconstrained multinomial MLE (black) is the most reliable reconstruction method\cite{de2008choice} in Fig \ref{fig:Figure_5} (a), and hence why we use it as the reference against which we calculate fidelity. However, while Gaussian MLE reconstructs states with higher fidelity to multinomial MLE than do the constrained and neural network methods, we still find some discrepancies in the two MLE methods.  Moreover, the two constrained MLE methods (green and red), and our neural network trained exclusively on pure states (magenta) behave similarly throughout the range of reconstruction scenarios, with their performance approximately tracking the purity of the state, as shown in the upper inset.  The close agreement between the neural network and the constrained MLE methods is somewhat surprising as the neural network is not constrained to reconstruct pure states; instead, it was merely not exposed to them during training.  This result points to the care needed to ensure that training sets reflect the system under investigation when using neural-network-based reconstruction methods.  }


{Additionally, we also include as an inset in Fig. \ref{fig:Figure_5} (a) the inference time on a GPU (RTX 2060) for estimating a density matrix with MLE constrained to pure states for the tomography data obtained from the IBM Q. 
For four qubits, MLE already takes several orders of magnitude longer for reconstruction than our neural network based system, as demonstrated in Fig. \ref{fig:Figure_2} (b).  
For comparison, we found that reconstruction with our system took approximately $0.8$ ms for four qubit systems.}

{Finally, we highlight an effect in which constrained reconstruction methods applied to systems with minor decoherence can overestimate the reconstruction fidelity when the programmed state is used as the reference state.  To emphasize this result, we consider only two-qubit systems and generate a test set of 20 random quantum states as previously described in ``Methods''.  We then initialize the qubits with the generated random quantum states and allow the system to self-optimize the quantum circuits\cite{shende2006synthesis} on a {seven qubit machine: $ibmq\_jakarta$ backend}. Note that we select the light optimization option in `qiskit.execute' from the Qiskit Terra API to balance between computational efficiency and accuracy of the transpiled circuit.  We then perform quantum state tomography and calculate the fidelity of the resulting states against the state programmed into the IBMQ, unlike in Fig. \ref{fig:Figure_5} (a) where multinomial MLE was used as the reference state. Fig. \ref{fig:Figure_5} (b) shows the resulting fidelity at each shot level with black, green, magenta, and orange corresponding to MLE with multinomial likelihood, MLE with multinomial likelihood constrained to only pure states, a neural network trained with ideal tomography data (pure states only), and a neural network trained with simulated tomography at 15 shots (pure states only), respectively.  We see that the constrained method and both pure-state trained neural networks overestimate the system's fidelity compared to unconstrained MLE.  In effect, these systems remove evidence of system noise by reconstructing pure states.  Such an effect can lead to undue confidence in the performance of a system and should be taken into account whenever constrained MLE or neural networks with limited training sets are used for reconstruction.  }

\section{Conclusion}
{In conclusion, we have experimentally investigated the resource and computation time scaling of quantum state tomography using machine learning techniques for pure states.  We used our developed machine-learning-based quantum state tomography system to reconstruct random quantum states created and measured with the IBM Q quantum computer for up to four qubit systems.  We also implement MLE for state reconstruction to cross-validate the efficacy of the network performance.
We show that our neural network can achieve relatively high reconstruction fidelities on average, for example, above $0.99$ for two qubits.  However, we also find that our neural network reconstruction technique suffers from a higher variance than comparable MLE reconstruction techniques.
Given the limited quantum volume of the IBM Q, we also had the opportunity to test our neural-based-reconstruction method on mixed quantum states.  Interestingly, our neural network, which is not constrained to reconstruct only pure states but has instead been trained exclusively on them, behaves similarly to constrained MLE methods in terms of reconstruction fidelity.  More specifically, constrained MLE and our neural network reconstruct mixed states with a fidelity that approximately tracks the state's purity.  This result points to the influence of the training set on the performance of neural-network-based tomography systems, and that one must ensure that a system under investigation is reflective of the training set. } 

{Lastly, we reiterate that while a neural network's training time increases for quantum systems of increasing dimension, the network need only be trained a single time. Furthermore, once trained, the inference time when using the neural network in practice is negligible.  In other words, our work reveals a trade-off between a high upfront training time with relatively high reconstruction fidelity and negligible reconstruction time using machine-learning-based techniques against slower but more accurate MLE-based methods.  As such, our results provide evidence for machine learning as a tool for full quantum state tomography in near-term intermediate-scale quantum hardware where system size is too large for MLE to be practical and the characteristics of a state under investigation do not need to be known with extreme accuracy. }

\section*{Acknowledgment} 
The views and conclusions contained in this document are those of the authors and should not be interpreted as representing the official policies, either expressed or implied, of the Army Research Laboratory or the U.S. Government. The U.S. Government is authorized to reproduce and distribute reprints for Government purposes notwithstanding any copyright notation herein. Additionally, we acknowledge use of the IBM Q for this work. The views expressed are those of the authors and do not reflect the official policy or position of IBM Quantum.  We would like to thank Joseph Lukens at Oak Ridge National Laboratory, Hanhee Paik at IBM and Sean Huver at NVIDIA for helpful discussions.

\appendix
\section*{Appendix}
\subsection{Parameter Tuning}
\label{app:par}
\Figure[h!](topskip=0pt, botskip=10pt, midskip=0pt)[width=.79\linewidth]{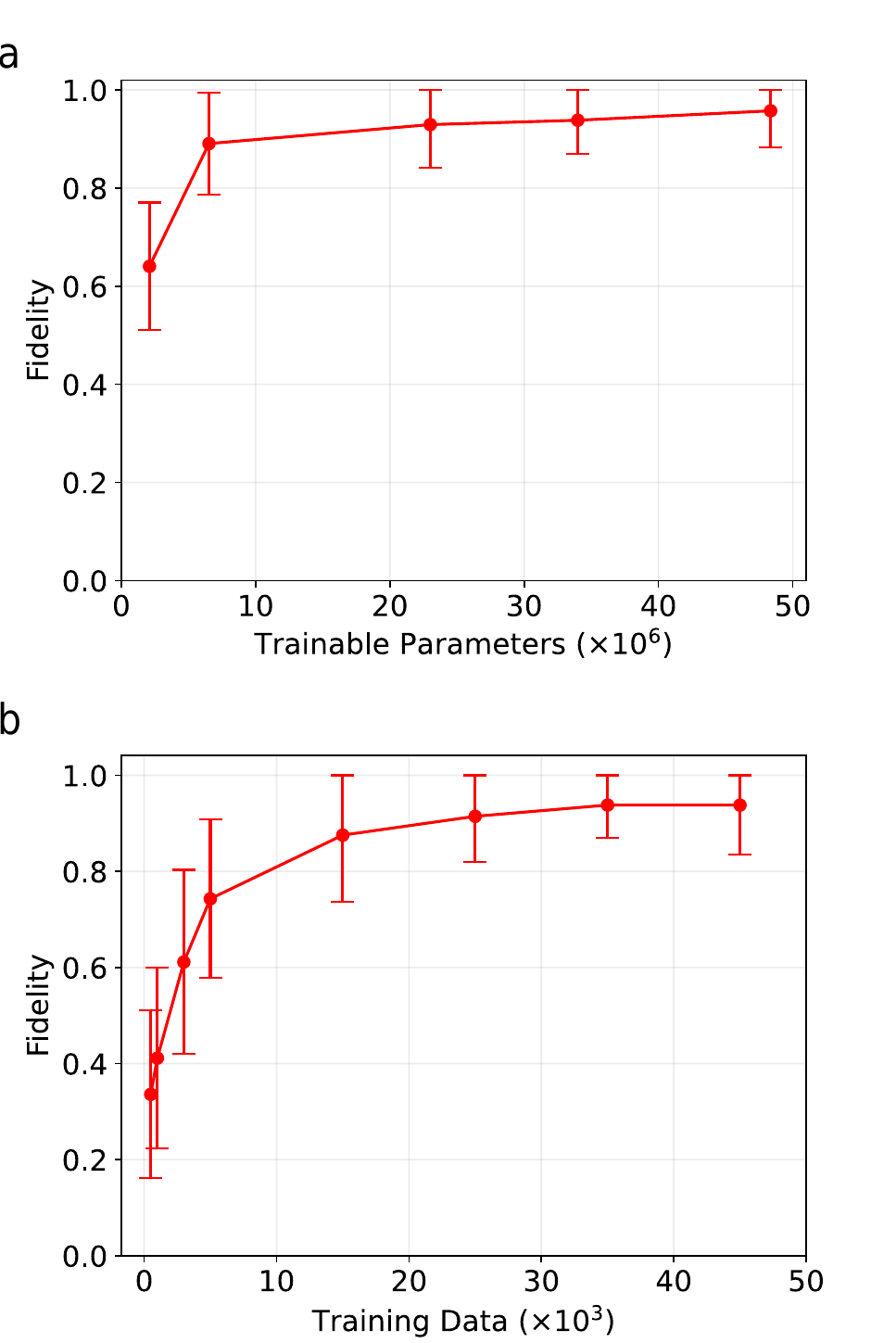}
{{\textbf{Examples of parameter tuning in the 4 qubit case.} Fidelity versus (a) trainable parameters and (b) size of the training set used to train the network. {The error bars in each figure represent one standard deviation from the mean.}}
\label{fig:appendix-1}}

{Neural network performance was optimized through manual tuning of the hyperparameters.
{This appendix illustrates the dependence of reconstruction fidelity on tuneable hyperparameters in the four-qubit case.}
First, we fix the number of quantum states in a training set at 35,000 and vary the network parameters by changing the number of neurons in the dense layers. We find that the fidelity increases with increasing trainable parameters and begins to saturate as shown in Fig. \ref{fig:appendix-1}(a). 
{Next, we set the trainable parameters at $33.99\times10^6$ and vary the number of quantum states in a training set.}
As expected, we find an increasing trend of fidelity with the size of the training set, as shown in Fig. \ref{fig:appendix-1}(b).
However, the fidelity starts to saturate after the training reaches 35,000.
{We note the actual hyperparameter choices used to generate the results used within the manuscript in Table \ref{table:one} and the inset of Fig. \ref{fig:Figure_2} (b).  We made the final hyperparameter selections based on the trade-offs demonstrated in this appendix.}
}

\subsection{Simulating the low-shot regime}
\label{app:low}
{Here we discuss how to simulate tomography in the {low-shot regime}. To illustrate the concept, we take the two-qubit case as an example.}

{Arrange the set of projectors $\{\mathcal{\hat{\varPi}}_i\}$ for $i\,=\,0,\,1,\, 2,\, ...,\, 35$, as discussed in Section II-A, in an array of shape $(9,\, 4,\, 4,\, 4)$, where the first axis (axis-0) represents Pauli bases in $\{X,Y,Z\}_{1}\otimes\{X,Y,Z\}_{2}$ and the second axis (axis-1) shows a set of $4\times4$ unitary matrices corresponding to eigenvectors of the given Pauli basis.}

\newcommand\mycommfont[1]{\footnotesize\ttfamily\textcolor{blue}{#1}}
\SetCommentSty{mycommfont}
\SetKwInput{KwInput}{Input}                
\SetKwInput{KwOutput}{Output}              
\DontPrintSemicolon
\begin{algorithm}[h]  
    \KwInput{A random quantum state ($\rho$)}
    \KwOutput{Measurements at shot `$S$'}
    Assume $\varPi$ represents the reshaped array of size of $(9,\, 4,\, 4,\, 4)$;\\
    \tcp {Create an empty list.}
    meas = [];\\ 
    \For{s $\leftarrow$ 0 to S-1}
    {
        \For{ j $\leftarrow$ 0 to 8}
                {
                    \tcp{$\varPi[.]$: Indexing along axis-0.}
                    
                    $\bar{n} \leftarrow map$\Big(Tr($\rho \varPi[j])$\Big)\\
                    \tcp{Create a mutable (36, ) array filled with zero.}
                    $z \leftarrow$ $zeros(36)$ \\
                    $1\leftarrow$ $z\Big[4\times j + argmax\big(multinomial(1, \bar{n})\big)\Big]$\\ 
                    meas$\leftarrow append\,(z)$
                }
    }
    meas $\leftarrow$ $reshape\,$\big(meas, $(1,\, 9\times S,\, 36)$\big)\\
    \textbf{return} $sum\,$(meas, $axis\,=\,1$)
            
\caption{Simulating tomography at shot `S'}
\end{algorithm}

\bibliographystyle{ieeetr}
\bibliography{IBMQ}

\EOD

\end{document}